\documentclass[runningheads]{llncs}
\usepackage[T1]{fontenc}
\usepackage{graphicx}
\usepackage{graphicx}
\usepackage{amsmath}
\usepackage{amssymb}
\usepackage{booktabs}
\usepackage{pifont}

\usepackage{bbding}
\usepackage{soul}
\usepackage{bibunits}
\usepackage{bm}
\usepackage{array}
\usepackage[dvipsnames]{xcolor} 
\usepackage{pdflscape}
\usepackage{makecell}
\usepackage{framed,multirow}
\usepackage{threeparttable}
\usepackage{amssymb}
\usepackage{pifont}
\usepackage{algorithm}
\usepackage{listings}
\usepackage{pythonhighlight}
\usepackage{float}
\usepackage{mathtools}
\usepackage{cuted}
\DeclareUnicodeCharacter{2212}{-}
\makeatletter
\newcommand\footnoteref[1]{\protected@xdef\@thefnmark{\ref{#1}}\@footnotemark}
\makeatother

\newcolumntype{P}[1]{>{\centering\arraybackslash}p{#1}}
\newlength\savewidth
\def\arrvline{\hfil\kern\arraycolsep\vline\kern-\arraycolsep\hfilneg}
\definecolor{mygray}{gray}{.9}
\definecolor{Highlight}{HTML}{39b54a}  

\newcolumntype{x}[1]{>{\centering\arraybackslash}p{#1pt}}
\newcolumntype{z}[1]{>{\raggedright\arraybackslash}p{#1pt}}

\definecolor{citecolor}{HTML}{0071BC}
\definecolor{linkcolor}{HTML}{ED1C24}
\usepackage[pagebackref,breaklinks,colorlinks,citecolor=citecolor]{hyperref}

\usepackage[T1]{fontenc}

\newcommand{\ourmethod}{{\fontfamily{ppl}\selectfont FSMNet}}
\begin{document}
%

\title{Accelerated Multi-Contrast MRI Reconstruction via Frequency and Spatial Mutual Learning}

\titlerunning{Accelerated MCMR via Frequency and Spatial Mutual Learning}


\author{Qi Chen\inst{1} \and 
Xiaohan Xing \inst{2,}\thanks{Corresponding author: Xiaohan Xing (\href{xhxing@stanford.edu}{xhxing@stanford.edu})}  \and 
Zhen Chen \inst{3} \and 
Zhiwei Xiong \inst{1}
 }


\institute{University of Science and Technology of China \and
Stanford University \and
 Centre for Artificial Intelligence and Robotics (CAIR), HKISI-CAS
}

\authorrunning{Qi Chen et al.}

%
\maketitle              
\begin{abstract}

To accelerate Magnetic Resonance (MR) imaging procedures, Multi-Contrast MR Reconstruction (MCMR) has become a prevalent trend that utilizes an easily obtainable modality as an auxiliary to support high-quality reconstruction of the target modality with under-sampled k-space measurements. The exploration of global dependency and complementary information across different modalities is essential for MCMR. However, existing methods either struggle to capture global dependency due to the limited receptive field or suffer from quadratic computational complexity. To tackle this dilemma, we propose a novel \underline{F}requency and \underline{S}patial \underline{M}utual Learning \underline{Net}work (\ourmethod), which efficiently explores global dependencies across different modalities. Specifically, the features for each modality are extracted by the \textit{Frequency-Spatial Feature Extraction} (FSFE) module, featuring a frequency branch and a spatial branch. Benefiting from the global property of the Fourier transform, the frequency branch can efficiently capture global dependency with an image-size receptive field, while the spatial branch can extract local features. To exploit complementary information from the auxiliary modality, we propose a \textit{Cross-Modal Selective fusion} (CMS-fusion) module that selectively incorporate the frequency and spatial features from the auxiliary modality to enhance the corresponding branch of the target modality. To further integrate the enhanced global features from the frequency branch and the enhanced local features from the spatial branch, we develop a \textit{Frequency-Spatial fusion} (FS-fusion) module, resulting in a comprehensive feature representation for the target modality. Extensive experiments on the BraTS and fastMRI datasets demonstrate that the proposed \ourmethod\ achieves state-of-the-art performance for the MCMR task with different acceleration factors. The code is available at: \url{https://github.com/qic999/FSMNet}.

\end{abstract}

\setcounter{footnote}{0}

\section{Introduction}\label{sec:introduction}
Accelerated Magnetic Resonance (MR) imaging, which reconstructs MR images from under-sampled k-space measurements, can significantly reduce the cost of the MR imaging procedure and improve the patient experience~\cite{tsao2012mri,han2019k,du2021adaptive}. 
However, the aliasing artifacts resulting from insufficient sampling often degrade the image quality and compromise clinical diagnoses~\cite{zhuo2006mr}. 
In clinical practice, multi-contrast MR images, such as T1 and T2 weighted images (T1WIs and T2WIs), are often simultaneously acquired to provide complementary structural information for diagnosis and treatment planning.
Recently, multi-contrast MR reconstruction (MCMR)~\cite{zhou2020dudornet,lyu2022dudocaf} has become a prevalent trend that utilizes an easily obtainable modality (e.g., T1WIs) as an auxiliary to support high-quality reconstruction of under-sampled target modalities (e.g., T2WIs). 

For multi-contrast MR images, similar structural features are distributed across different regions within each modality, and complementary information exists across different modalities. 
Therefore, the inherent problem of the MCMR task is how to comprehensively explore long-range dependencies within each modality and sufficiently leverage complementary information from the auxiliary modality.
Toward this goal, a series of methods \cite{feng2021multi,li2023multi,xing2024comprehensive,xing2023gradient,xing2022discrepancy} integrate multi-modal features based on convolutional neural networks (CNNs). However, the actual receptive field of CNNs is restricted due to the vanishing gradient issue, thus failing to capture global dependencies.
Recent advances in vision transformers \cite{dosovitskiy2020image} have led to self-attention-based feature integration methods \cite{feng2022multi,huang2023accurate}, enabling the modeling of long-range dependencies and enhancing the performance of the MCMR task. However, these methods \cite{feng2022multi,huang2023accurate} face significant challenges in clinical deployment due to their quadratic computation complexity ($\mathcal{O}(N^2)$ for $N$ tokens).
Therefore, there is highly demanded to develop an efficient method that can capture global dependencies and extract complementary information from both modalities without introducing a heavy computational burden.

Inspired by \cite{mao2021deep,zhou2023fourmer}, we resort to frequency information in the Fourier domain as an efficient global feature extractor.
Since each pixel in the Fourier domain interacts with all pixels in the spatial domain, the frequency features naturally encompass global properties and can achieve an image-size receptive field.
Thus, the global features extracted in the Fourier domain complement the local features extracted in the spatial domain, providing a more efficient way for global feature interaction across different modalities.
In light of this, reconstruction accuracy for the target modality can be improved by addressing the following issues:
1) How to enhance the feature representation of each modality with the assistance of the global view from the Fourier domain.
2) How to effectively integrate the global frequency and local spatial features across different modalities.

In this work, we propose a novel \underline{F}requency and \underline{S}patial \underline{M}utual learning \underline{Net}work (\ourmethod) for MCMR, offering an efficient method to extract and integrate global and local features across different modalities.
\ourmethod\ extracts features for each modality using a stack of \textit{Frequency-Spatial Feature Extraction} (FSFE) modules, which include a frequency branch and a spatial branch. The frequency branch is designed to capture global features with an image-size receptive field, while the spatial branch specializes in extracting local structural features.
To fully incorporate complementary information from the auxiliary modality, we introduce a \textit{Cross-Modal Selective fusion} (CMS-fusion) module. This module selectively integrates frequency and spatial features from the auxiliary modality to enhance the respective branch of the target modality.
To further integrate enhanced global features from the frequency branch and enhanced local features from the spatial branch, we present a \textit{Frequency-Spatial fusion} (FS-fusion) module that exchanges beneficial information across these two branches based on an adaptive attention map.
The main contributions are summarized as:
\begin{itemize}
    \item Our FSMNet provides an efficient method for capturing global dependency via Fourier transform. The proposed FSFE module enhances feature representations for each modality by extracting global features from the frequency branch and local features from the spatial branch. (\S\ref{sec:FSFE})
    \item We propose a CMS-fusion module to enhance global and local features by selectively incorporating complementary information from the auxiliary modality, along with an FS-fusion module to further integrate the enhanced global and local features for the target modality. (\S\ref{sec:feature_fusion})
    \item Experimental results on two MRI datasets (BraTS~\cite{menze2014multimodal} and fastMRI~\cite{zbontar2018fastmri}) demonstrate that \ourmethod\ outperforms existing MCMR methods across various acceleration factors. (\S\ref{sec:experimental_results})
\end{itemize}

\begin{figure}[t]
	\centering
        \setlength{\belowcaptionskip}{-0.3cm}
	\includegraphics[width=0.95\columnwidth]{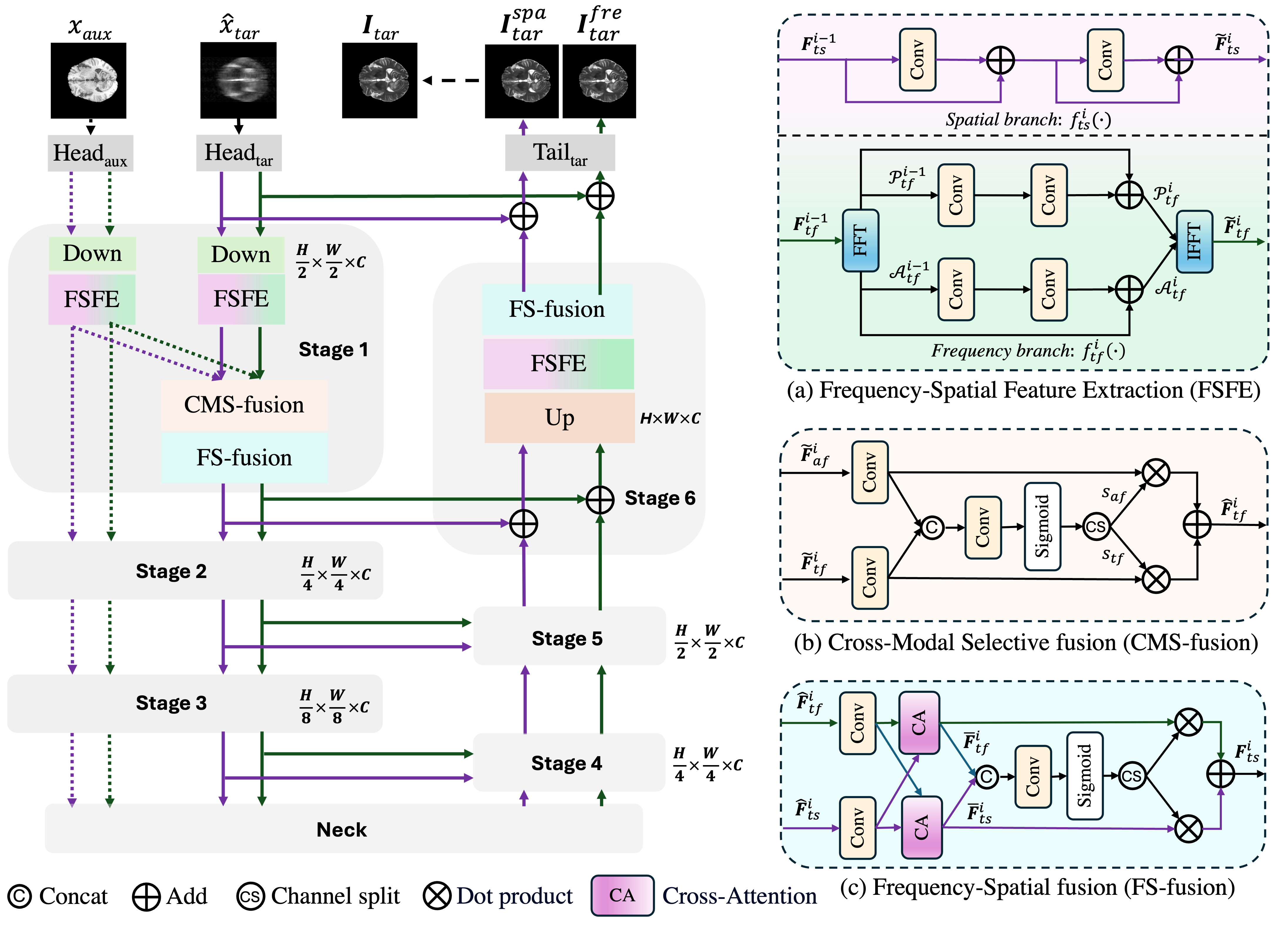}
 \caption{
    \textbf{Overview of \ourmethod}: In each stage, the FSFE module extracts global and local features from the frequency and spatial branches, respectively. The CMS-fusion module integrates the multi-modal features for each branch, and subsequently, the FS-fusion module combines the features across the frequency and spatial branches.
    }
	\label{fig:method}
\end{figure}

\section{Methodology}\label{sec:method}
An overview of \ourmethod\ is depicted in Fig.~\ref{fig:method}. Given the under-sampled target image $\hat{x}_{tar}$ and the fully-sampled image $x_{aux}$ from the auxiliary modality, \ourmethod\ aims to reconstruct a high-quality image $\boldsymbol{I}_{tar}$ for the target modality.
Each modality comprises two branches, the frequency branch (green lines) and the spatial branch (purple lines). In each encoder stage (Stage 1-3), the FSFE module extracts global features from the frequency branch and local features from the spatial branch. A total of 4 sets of features are encoded for the two modalities. The CMS-fusion module selectively integrates features from both modalities to generate enhanced global and local features for the target modality, which are mutually enhanced by the subsequent FS-fusion module.
The neck stage mirrors the encoder stages but omits the downsample operation. As only target modality is decoded, the decoder stage (Stage 4-6) consist of the upsample operation, FSFE module, and FS-fusion module. The image $\boldsymbol{I}_{tar}^{spa}$ recovered by the spatial branch is considered the high-quality image $\boldsymbol{I}_{tar}$ for the target modality.

\subsection{Frequency-Spatial Feature Extraction}\label{sec:FSFE}
Since similar structural features are distributed across different regions in an MR image, improving MRI reconstruction performance entails capturing global dependency and integrating similar features from various image locations. To address the limitation of CNNs in global modeling, we propose the \textit{Frequency-Spatial Feature Extraction} (FSFE) module. Benefiting from the global property of Fourier transform, the FSFE module efficiently extracts local and global features through the spatial and frequency branches for both modalities.

Taking the $i$-th stage of the target modality as an example, the input spatial and frequency features (i.e., $\boldsymbol{F}_{ts}^{i-1}$ and $\boldsymbol{F}_{tf}^{i-1}$ obtained from the previous stage) are fed into the spatial and frequency branches, respectively,
\begin{equation}
\begin{split}
    \widetilde{\boldsymbol{F}}_{ts}^{i}=f_{ts}^{i}(\boldsymbol{F}_{ts}^{i-1}), 
    \widetilde{\boldsymbol{F}}_{tf}^{i}=f_{tf}^{i}(\boldsymbol{F}_{tf}^{i-1}),  
\end{split}
\end{equation}
where $f_{ts}^i(\cdot)$ and $f_{tf}^i(\cdot)$ represent the networks for the spatial and frequency branches at the $i$-th stage for the target modality. 
As illustrated in Fig. \ref{fig:method} (a), the spatial branch consists of a cascade of residual convolutional blocks~\cite{he2016deep}, extracting local features within the limited receptive field. 
In the frequency branch, the input feature $\boldsymbol{F}_{tf}^{i-1}$ is first converted to the frequency domain via Fast Fourier Transform (FFT)~\cite{elliott1982fast}. Subsequently, the frequency feature is decomposed into amplitude and phase components (i.e., $\mathcal{A}_{tf}^{i-1}$, $\mathcal{P}_{tf}^{i-1}$), defined as:
\begin{equation}
    \begin{aligned}
    & \mathcal{A}_{tf}^{i-1}=\sqrt{\mathcal{R}^2(\mathrm{FFT}(\boldsymbol{F}_{tf}^{i-1}))+\mathcal{I}^2(\mathrm{FFT}(\boldsymbol{F}_{tf}^{i-1}))}, \\
    & \mathcal{P}_{tf}^{i-1}=\arctan \left[\mathcal{I}(\mathrm{FFT}(\boldsymbol{F}_{tf}^{i-1})) / \mathcal{R}(\mathrm{FFT}(\boldsymbol{F}_{tf}^{i-1}))\right],
\end{aligned}
\end{equation}
where $\mathcal{R}(\cdot)$ and $\mathcal{I}(\cdot)$ represent the real and imaginary parts of the frequency feature.
According to Fourier transform theory~\cite{nussbaumer1982fast}, each pixel in the Fourier domain interacts with all pixels in the spatial domain, enabling image-size receptive field and global feature extraction by extracting features from the amplitude and phase components in the Fourier domain.
As depicted in Fig.~\ref{fig:method} (a), we employ two convolutional layers with a residual connection to extract features $\mathcal{A}_{tf}^{i}$ and $\mathcal{P}_{tf}^{i}$ from the two components, respectively.
The output $\widetilde{\boldsymbol{F}}_{tf}^{i}$ of the frequency branch can be obtained via inverse Fourier transform (IFFT).
Due to the global property of the Fourier transform, $\widetilde{\boldsymbol{F}}_{tf}^{i}$ contains global information from the target modality, which is complementary to the local feature $\widetilde{\boldsymbol{F}}_{ts}^{i}$.

Similarly, given the input features $\boldsymbol{F}_{as}^{i-1}$ and $\boldsymbol{F}_{af}^{i-1}$ for the $i$-th stage of the auxiliary modality, the FSFE module extracts local and global features $\widetilde{\boldsymbol{F}}_{as}^{i}$ and $\widetilde{\boldsymbol{F}}_{af}^{i}$ from the spatial and frequency branches, respectively.

\subsection{Cross-Modal Frequency-Spatial Feature Fusion}\label{sec:feature_fusion}
For multi-contrast MR images, local features (i.e., $\widetilde{\boldsymbol{F}}_{as}^{i}$ and $\widetilde{\boldsymbol{F}}_{ts}^{i}$) and global features (i.e., $\widetilde{\boldsymbol{F}}_{af}^{i}$ and $\widetilde{\boldsymbol{F}}_{tf}^{i}$) are extracted by the spatial and frequency branches, respectively.
Given the existence of complementary information across different modalities, as well as the spatial and frequency branches, there is a pressing need to design an effective fusion strategy to explore this complementary information. Toward this goal, we first fuse the multi-modal features for the different branches separately, and then integrate the spatial and frequency branches to generate a comprehensive feature representation for the target modality.

\smallskip\noindent\textbf{Cross-Modal Selective fusion (CMS-fusion):}
The CMS-fusion module generates enhanced local and global features by selectively integrating features from the respective branch in the auxiliary modality.
Taking the frequency branch as an example, the features $\widetilde{\boldsymbol{F}}_{tf}^{i}$ and $\widetilde{\boldsymbol{F}}_{af}^{i}$ are processed by two convolutional layers, then combined to learn the pixel-wise modality complementary score $s_{tf}$ and $s_{af}$ for each modality: 

\begin{equation}
[s_{tf}, s_{af}]=\operatorname{sigmoid}(\operatorname{conv}[\widetilde{\boldsymbol{F}}_{tf}^{i}, \widetilde{\boldsymbol{F}}_{af}^{i}]).
\label{channel_attn}
\end{equation}
The complementary features from different modalities are fused to generate an enhanced global feature $\widehat{\boldsymbol{F}}_{tf}^{i}$ for the target modality:
\begin{equation}
\widehat{\boldsymbol{F}}_{tf}^{i} = 
s_{tf} * \widetilde{\boldsymbol{F}}_{tf}^{i} + s_{af} * \widetilde{\boldsymbol{F}}_{af}^{i}.
\label{cms_fusion}
\end{equation}

Similarly, the local features $\widetilde{\boldsymbol{F}}_{as}^{i}$ and $\widetilde{\boldsymbol{F}}_{ts}^{i}$ from the spatial branch are selectively integrated via the CMS-fusion module, producing the enhanced local feature $\widehat{\boldsymbol{F}}_{ts}^{i}$ for the target modality.
CMS-fusion in the spatial branch enables the exploration of detailed structural information from the auxiliary modality.

\smallskip\noindent\textbf{Frequency-Spatial fusion (FS-fusion):}

By incorporating complementary information from the auxiliary modality, the enhanced features $\widehat{\boldsymbol{F}}_{tf}^{i}$ and $\widehat{\boldsymbol{F}}_{ts}^{i}$ offer a better representation of global information (corresponding to the basic structure and overall appearance) and local information (such as detailed structures), respectively. 
Given that global and local information are complementary and both are crucial for MRI reconstruction, we introduce the FS-fusion module to adaptively merge the global frequency information and local spatial information.

As shown in Fig. \ref{fig:method} (c), the FS-fusion module employs the cross-attention mechanism to achieve mutual enhancement of the spatial and frequency branches.
Aiming at enhancing the spatial information by exploring complementary frequency information, 
$\widehat{\boldsymbol{F}}_{ts}^{i}$ is used as the query (\textit{Q}), and $\widehat{\boldsymbol{F}}_{tf}^{i}$ serves as both the key (\textit{K}) and value (\textit{V}). The integration process can be formalized as follows:
\setlength\abovedisplayskip{4pt}
\setlength\belowdisplayskip{4pt}
\begin{equation}
    \overline{\boldsymbol{F}}_{ts}^{i} = f(\text{softmax}\left(\widehat{\boldsymbol{F}}_{tf}^{i} ({\widehat{\boldsymbol{F}}_{ts}^{i}})^T / \sqrt{d}\right) \widehat{\boldsymbol{F}}_{ts}^{i}) + \widehat{\boldsymbol{F}}_{ts}^{i},
\end{equation}

where $f$ is the convolution function and $\overline{\boldsymbol{F}}_{ts}^{i}$ is the spatial feature enhanced by the global frequency information. Similarly, we obtain the enhanced frequency feature $\overline{\boldsymbol{F}}_{tf}^{i}$. 
The enhanced features $\overline{\boldsymbol{F}}_{ts}^{i}$ and $\overline{\boldsymbol{F}}_{tf}^{i}$ are integrated into $\boldsymbol{F}_{ts}^{i}$ using pixel-wise selective fusion, as described in Eq. \eqref{channel_attn} and Eq. \eqref{cms_fusion}.

\subsection{Loss Function}\label{sec:LF}
FSMNet reconstructs target modality image $\boldsymbol{I}_{tar}^{spa}$ and $\boldsymbol{I}_{tar}^{fre}$ from the spatial and frequency branch, respectively. 
The model is supervised by pixel-level loss $\mathcal{L}_{\mathrm{pixel}}$ and frequency-level loss $\mathcal{L}_{\mathrm{fre}}$, defined as:
\vspace{-1mm}
\begin{small}
    \begin{equation}
    \mathcal{L}_{\mathrm{pixel}}=||\boldsymbol{I}_{tar}^{spa}-\boldsymbol{I}_{tar}^{full}||_1+||\boldsymbol{I}_{tar}^{fre}-\boldsymbol{I}_{tar}^{full}||_1, \\
    \end{equation}
    \begin{equation}
    \mathcal{L}_{\mathrm{fre}}=||\mathcal{A}(\mathrm{FFT}(\boldsymbol{I}_{tar}^{fre}))-\mathcal{A}( \mathrm{FFT}(\boldsymbol{I}_{tar}^{full}))||_1+||\mathcal{P}(\mathrm{FFT}(\boldsymbol{I}_{tar}^{fre}))-\mathcal{P}(\mathrm{FFT}(\boldsymbol{I}_{tar}^{full}))||_1,
    \end{equation}
\end{small}
where $\boldsymbol{I}_{tar}^{full}$ corresponds to the fully-sampled target modality image, serving as the ground-truth. $\mathcal{A}(\cdot)$ and $\mathcal{P}(\cdot)$ denote the amplitude and phase components in the Fourier domain.
The total loss is defined as:
\setlength\abovedisplayskip{3pt}
\setlength\belowdisplayskip{3pt}
\begin{equation}
    \mathcal{L} = \mathcal{L}_{\mathrm{pixel}} + \lambda * \mathcal{L}_{\mathrm{fre}},
\end{equation}
where $\lambda$ is the trade-off coefficient and is empirically set as $0.01$.

\section{Experiments}\label{sec:experiments}

\begin{table*}[t]
    \centering
    \footnotesize
    \caption{
    \textbf{Quantitative results on the BraTS and fastMRI datasets with different acceleration factors.} We report mean±std for the PSNR and SSIM metrics. 
    }
    \scriptsize
    \begin{tabular}{p{0.18\linewidth}|p{0.14\linewidth}|P{0.14\linewidth}P{0.14\linewidth}|P{0.14\linewidth}P{0.14\linewidth}}
    \toprule
     \textbf{BraTS} & &\multicolumn{2}{c|}{$4 \times$} & \multicolumn{2}{c}{$8 \times$} \\
     Method & Year & PSNR & SSIM & PSNR & SSIM \\ \midrule
     Zero-filling~\cite{bernstein2001effect} & JMRI'01 & 30.04+1.51 &0.748±0.034  & 26.58±1.49 & 0.673±0.036 \\
     MDUNet~\cite{xiang2018deep} & TBME'18 &37.94±1.66 &0.975±0.006 &35.19±1.64 &0.960±0.009 \\
     MINet~\cite{feng2021multi} & MICCAI'21 & 38.26±1.74 &0.976±0.006 &35.23±1.72 &0.961±0.009 \\
     MCCA~\cite{li2023multi} & JBHI'23 & 38.03±1.68 &0.975±0.006 &35.37±1.66 &0.962±0.009 \\
     SwinIR~\cite{liang2021swinir} & CVPR'21 & 37.87±1.73 &0.974±0.006 &34.95±1.72 &0.960±0.009\\
     MTrans~\cite{feng2022multi} & TMI'22 & 36.02±1.67 &0.962±0.007 &34.81±1.57 &0.957±0.009 \\
     DCAMSR~\cite{huang2023accurate} & MICCAI'23 & 38.60±1.75 &0.978±0.006 &35.99±1.74 &0.965±0.009 \\
     Ours (\ourmethod) & - &\textbf{41.76}±1.88 &\textbf{0.986}±0.004 &\textbf{38.60}±1.85 &\textbf{0.977}±0.006 \\
     \toprule
    \textbf{fastMRI} & &\multicolumn{2}{c|}{$4 \times$} & \multicolumn{2}{c}{$8 \times$} \\
     Method & Year & PSNR & SSIM & PSNR & SSIM \\
    \midrule
     Zero-filling~\cite{bernstein2001effect} & JMRI'01 & 24.5±1.37 &0.442±0.10  & 22.9±1.25 & 0.369±0.12\\
     MDUNet~\cite{xiang2018deep} & TBME'18 &28.6±1.00 &0.600±0.05 &27.9±0.86 &0.544±0.05 \\
     MINet~\cite{feng2021multi} & MICCAI'21 &29.4±1.88 &0.639±0.06 &28.1±1.74 &0.563±0.08\\
     MCCA~\cite{li2023multi} & JBHI'23 &29.4±1.87 &0.637±0.06 &28.2±1.75 &0.562±0.08 \\
     SwinIR~\cite{liang2021swinir} & CVPR'21 &29.4±1.87 &0.636±0.06 &28.1±1.74 &0.560±0.08 \\
     MTrans~\cite{feng2022multi} & TMI'22 &29.0±1.79 &0.619±0.06 &27.3±1.68 &0.526±0.08 \\
     DCAMSR~\cite{huang2023accurate} & MICCAI'23 &29.4±1.87 &0.637±0.06 &28.4±1.79 &0.569±0.08 \\
     Ours (\ourmethod) & - &\textbf{29.7}±1.93 &\textbf{0.646}±0.07 &\textbf{28.5}±1.80 &\textbf{0.572}±0.08 \\
    \bottomrule
    \end{tabular}
\label{tab:main_result}
\end{table*}

\subsection{Datasets and Implementation Details}

\noindent\textbf{Dataset description:} We evaluate our method on the BraTS~\cite{menze2014multimodal} and fastMRI~\cite{zbontar2018fastmri} datasets. For BraTS dataset, we use T1-weighted (T1WI) and T2-weighted (T2WI) brain MRI volumes, with T1WI supporting T2WI acceleration. We randomly select 100 patients, sample slices from the 3D volumes, and split the dataset 3:1 for training and testing. The fastMRI dataset consists of single-coil proton density-weighted (PDWI) and fat-suppressed proton density-weighted (FS-PDWI) knee MRI volumes. Following~\cite{feng2021multi}, we filter out 227 PDWI and FS-PDWI pairs for training and 24 pairs for testing, using PDWI as the auxiliary and FS-PDWI as the target modality. The reconstruction performance is evaluated under acceleration factors (AF) of 4× and 8× using the Peak Signal-to-Noise Ratio (PSNR) and Structural Similarity Index Measure (SSIM).

\noindent\textbf{Implementation details:} We implement \ourmethod\ and all comparison methods on RTX 3090 GPUs using the PyTorch library \cite{paszke2019pytorch}. The image resolution is $240\times 240$ and $320\times 320$ for the BraTS and fastMRI datasets, respectively. \ourmethod\ is trained using AdamW optimizer with $\beta_1=0.9$ and $\beta_2=0.999$. The training process for \ourmethod\ consists of 100k iterations with a batch size of 4. The learning rate is initially set to $1 \times {10}^{-4}$ and decays by a factor of 0.1 every 20k iterations. The comparison methods are implemented following the original configurations described in their respective papers \cite{xiang2018deep,feng2021multi,li2023multi,liang2021swinir,feng2022multi,huang2023accurate}. 

\subsection{Experimental Results}\label{sec:experimental_results}
\noindent\textbf{Comparison with state-of-the-art methods:} We conduct a comprehensive comparison between \ourmethod\ and state-of-the-art MCMR methods on both datasets. As shown in \tableautorefname~\ref{tab:main_result}, \ourmethod\ consistently outperforms existing MCMR methods \cite{xiang2018deep,feng2021multi,li2023multi,liang2021swinir,feng2022multi,huang2023accurate} across various acceleration factors on both the BraTS and fastMRI datasets.

\begin{figure}[t]
	\centering
        \setlength{\abovecaptionskip}{-0.05cm}
	\includegraphics[width=\columnwidth]{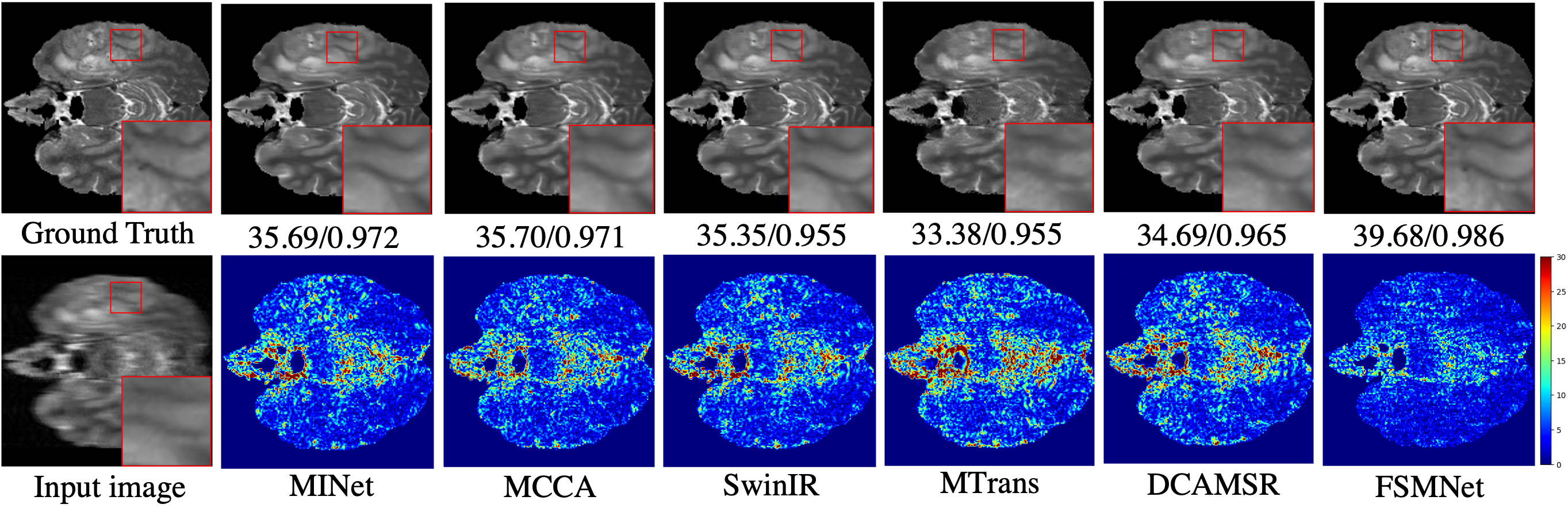}
 \caption{
    \textbf{Qualitative visualization} of the reconstructed images (1st row) and error maps (2nd row) for different MCMR methods with $4\times$ AF on the BraTS dataset. 
    Additional qualitative results are provided in the supplementary material. 
    } 
	\label{fig:qualitative_evaluation}
\end{figure}

Specifically, when compared with the current state-of-the-art method (i.e., DCAMSR \cite{huang2023accurate}), our method improves the PSNR by 3.16 dB and 2.61 dB under $4\times$ and $8\times$ AF on the BraTS dataset, respectively.
We further illustrate the reconstructed images and error maps of different methods in Fig.~\ref{fig:qualitative_evaluation}. 
For the example with $4\times$ AF on the BraTS dataset, \ourmethod\ recovers more image details and leads to a smaller error map. 
More examples are provided in the supplementary material.
The visualization results align with the improvement in the PSNR and SSIM metrics shown in Table \ref{tab:main_result}.

\begin{table*}[t]
    \centering
    \footnotesize
    \caption{
    \textbf{Ablation study on the BraTS dataset with different AF.} 
    }
    \scriptsize
    \begin{tabular}{P{0.12\linewidth}|P{0.12\linewidth}|P{0.12\linewidth}|P{0.13\linewidth}P{0.13\linewidth}|P{0.13\linewidth}P{0.13\linewidth}}
     \toprule
      \multirow{2}{*}{FSFE} & \multirow{2}{*}{CMS-fusion} & \multirow{2}{*}{FS-fusion} &\multicolumn{2}{c|}{$4 \times$} & \multicolumn{2}{c}{$8 \times$} \\
      & & & PSNR & SSIM & PSNR & SSIM \\ \midrule
      & & &37.94±1.66 &0.975±0.006 &35.19±1.64 &0.960±0.009 \\
     \Checkmark&  & &41.05±1.80  &0.985±0.004 & 37.97±1.78 & 0.974±0.007 \\
     \Checkmark&\Checkmark &  &41.39±1.83  &0.986±0.004 & 38.27±1.78 & 0.975±0.007 \\
     \Checkmark&\Checkmark  &\Checkmark  &41.76±1.88 &0.986±0.004 &38.60±1.85 &0.977±0.006 \\
    \bottomrule
    \end{tabular}
\label{tab:ablation}
\end{table*}

\noindent\textbf{Ablation Study:}
For the T2WI MRI reconstruction on the BraTS dataset with different AF, we conduct an ablation study to evaluate the contributions of our proposed FSFE, CMS-fusion, and FS-fusion modules. As shown in Table~\ref{tab:ablation}, compared with the vanilla variant (multi-modal UNet with concat fusion, $1st$ row), the reconstruction performance significantly improves with the introduction of the FSFE module ($2nd$ row). Specifically, the PSNR improves by 3.11 dB and 2.78 dB under $4 \times$ and $8 \times$ AF, respectively.
These results indicate that the global frequency and local spatial features extracted by the FSFE module can effectively enhance the feature representation and benefit MRI reconstruction.
As demonstrated in the 3-4 rows of Table~\ref{tab:ablation}, we observe performance degradation when replacing the CMS-fusion and FS-fusion with the simple summation fusion method. 
These results suggest that the proposed CMS-fusion and FS-fusion modules can effectively explore complementary information across different modalities and branches. Moreover, a feature analysis for the FS-fusion module is provided in the supplementary material.

\section{Conclusion}\label{sec:conclusion}
For the accelerated MCMR task, we propose a novel \ourmethod\ to effectively explore and integrate the global frequency feature and local spatial feature across different modalities.
In our proposed FSFE module, the spatial branch captures local dependency, while the frequency branch efficiently captures global dependency and achieves image-size receptive field via Fourier transform. 
The CMS-fusion is devised to generate enhanced local and global features by adaptively integrating complementary information from the auxiliary modality. 
The enhanced local and global features are further integrated via the FS-fusion module to produce a comprehensive feature representation for the target modality.
Extensive experiments on the BraTS and fastMRI datasets demonstrate the superiority of \ourmethod\ for the MCMR task under various acceleration factors.

\begin{credits}

\subsubsection{\discintname}
The authors have no competing interests to declare that are relevant to the content of this article.
\end{credits}

%
%
%
\bibliographystyle{splncs04}
\bibliography{refs}

\clearpage
\appendix

\end{document}


%

\title{Supplementary Material \\ Accelerated Multi-Contrast MRI Reconstruction via Frequency and Spatial Mutual Learning}

\titlerunning{Accelerated MCMR via Frequency and Spatial Mutual Learning}
\author{Qi Chen\inst{1} \and 
Xiaohan Xing \inst{2,}\thanks{Corresponding author: Xiaohan Xing (\href{xhxing@stanford.edu}{xhxing@stanford.edu})}  \and 
Zhen Chen \inst{3} \and 
Zhiwei Xiong \inst{1}
 }


\institute{University of Science and Technology of China \and
Stanford University \and
 Centre for Artificial Intelligence and Robotics (CAIR), HKISI-CAS
}

\authorrunning{Qi Chen et al.}


%
\maketitle              

\begin{figure}[h]
\centering
\includegraphics[width=0.8\columnwidth]{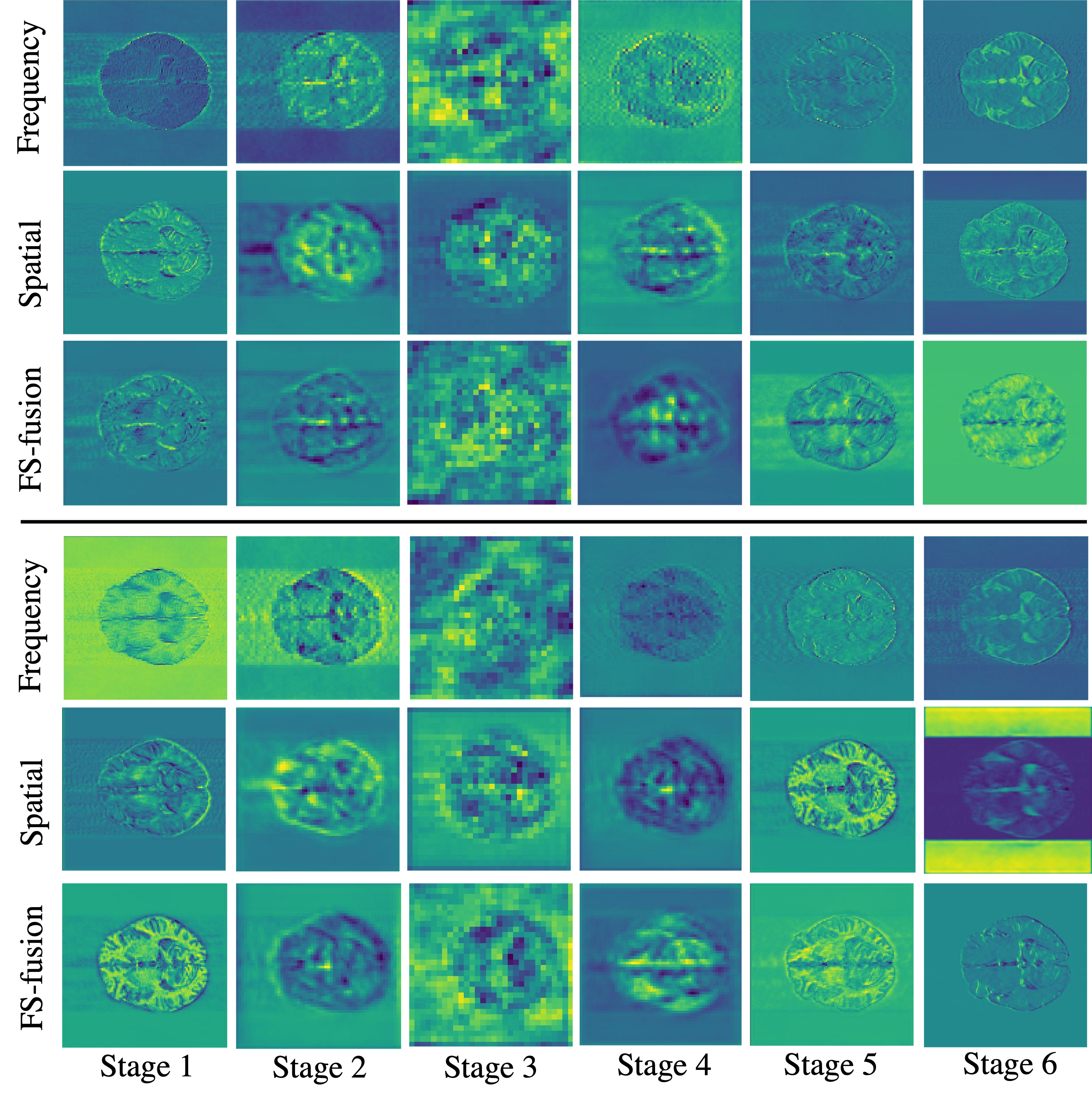}
\caption{Feature visualization. A feature analysis was conducted for the frequency and spatial features based on the FS-fusion module, which includes $\hat{\boldsymbol{F}}_{tf}^{i}, \hat{\boldsymbol{F}}_{ts}^{i}, {\boldsymbol{F}}_{ts}^{i}$. Two channels of these features were randomly selected from Stage1-6 for visualization. The results align with our motivation, whereby the frequency feature contains more global information, the spatial feature encompasses rich local features, and the final feature constitutes a comprehensive feature that is enhanced by both the frequency and spatial features.} 
\label{fig:supp_feature_visualization}
\end{figure}

\begin{figure}[t]
	\centering
	\includegraphics[width=\columnwidth]{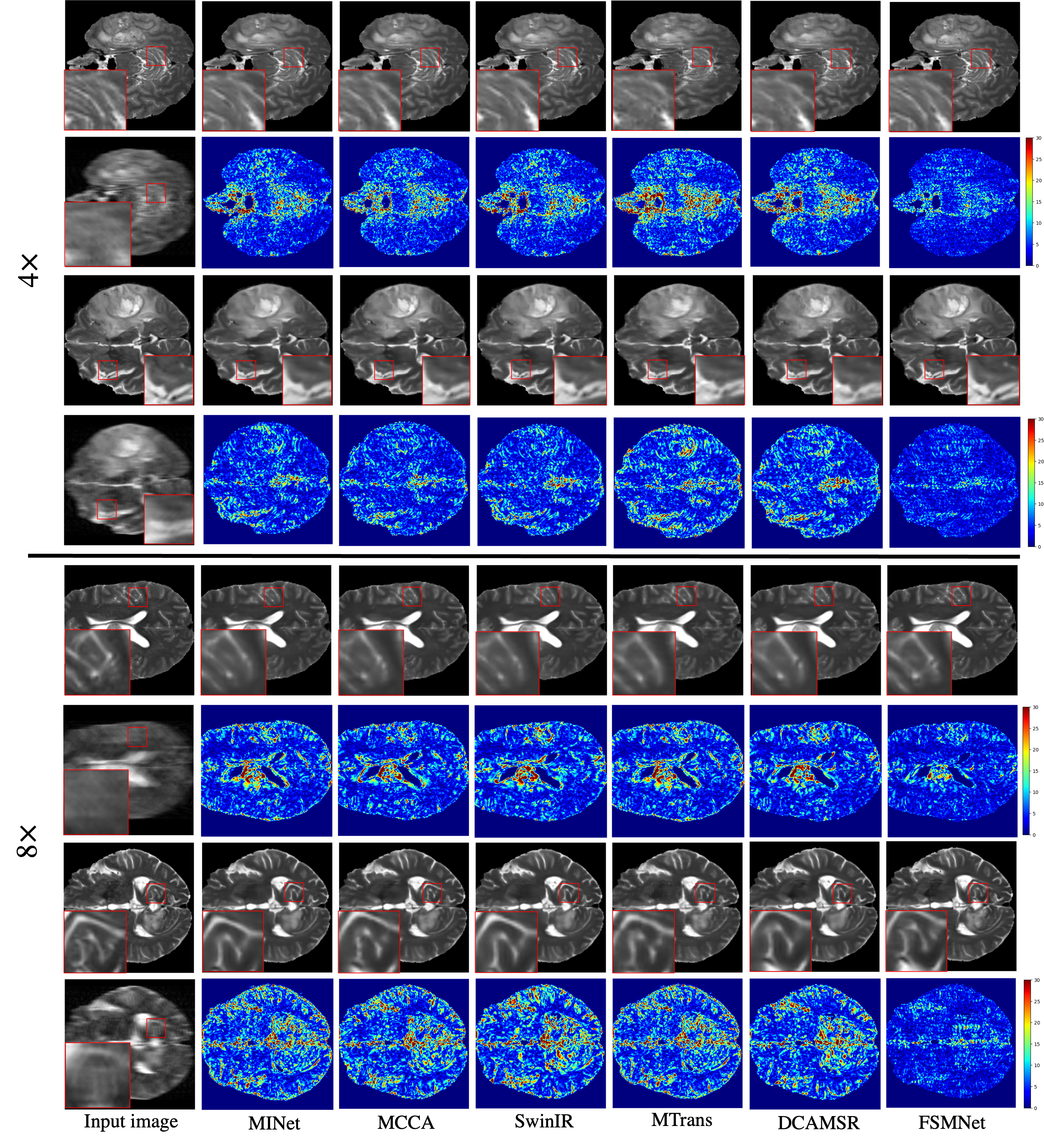}
 \caption{
    More qualitative visualizations compared to different MCMR methods with $4\times$ and $8\times$ AF on the BraTS dataset. 
    } 
	\label{fig:supp_qualitative_results}
\end{figure}